\documentclass[a4paper,12pt]{article}
\usepackage{amsmath}
\usepackage{pstricks}
\usepackage{pst-node}
\usepackage[ansinew]{inputenc}
\usepackage{amssymb,amsmath}
\usepackage{amsfonts}
\usepackage{epsfig}

\def\p{\partial}
\def\px{\partial_x}
\def\py{\partial_y}
\def\a{\alpha}
\def\b{\beta}
\def\g{\gamma}
\def\d{\delta}
\def\o{\omega}

\def\vp{\varphi}

\font\Sets=msbm10

\def\Complex {\hbox{\Sets C}}    \def\Rational {\hbox{\Sets Q}}

\def\be{\begin{equation}}       \def\ba{\begin{array}}

\def\ee{\end{equation}}         \def\ea{\end{array}}

\def\bea {\begin{eqnarray}}      \def\eea {\end{eqnarray}}

\def\bean{\begin{eqnarray*}}    \def\eean{\end{eqnarray*}}

\def\pa  {\partial}

\def\const {\mathop{\rm const}\nolimits}

\def\qed   {\vrule height0.6em width0.3em depth0pt}

\def\<{\langle} \def\({\left(}  \def\>{\rangle} \def\){\right)}

\newtheorem{exi}{Example}

\author{E. Kartashova  \\\\
 RISC, J.Kepler University, Linz, Austria\\\\
e-mail:
lena@risc.uni-linz.ac.at }

\title{Hierarchy of general invariants for bivariate LPDOs.}

\begin{document}
\date{}
\maketitle

\abstract We study invariants under gauge transformations of
linear partial differential operators  on two variables. Using
results of BK-factorization, we construct hierarchy of general
invariants for operators of an arbitrary order. Properties of
general invariants are studied and some examples are presented. We
also show  that classical Laplace invariants correspond to some
particular cases
of general invariants.\\


\section{Introduction}
Laplace invariants $\hat{a}= c- ab -a_x \quad \mbox{and} \quad
\hat{b}=c- ab  -b_y$ , introduced for  bivariate hyperbolic
operator of the second order

\be \label{Lap}\p_x \p_y + a\p_x + b\p_y + c, \ee

are quite well-known and have been studied by many researchers
(see, for instance, \cite{sha1995} and bibl. herein). Their
importance is due to the classical theorem:
\paragraph{Theorem 1}
Two operators of the form
 are equivalent under gauge transformations {\bf iff}
when their Laplace invariants coincide
pairwise.\\

 These  invariants are used extensively
in the theory of integrability (\cite{sh1981}, \cite{lez},
\cite{bianchi}, \cite{ziz}, \cite{tsarev96},\cite{tsarev98} and
others).  Obviously, Laplace invariants can be regarded as
factorization's "remainders" for the initial operator (\ref{Lap}):

\begin{equation}\label{dar}
\p_x \p_y + a\p_x + b\p_y + c = \left\{\begin{array}{c}
(\p_x + b)(\p_y + a) - ab - a_x + c\\
(\p_y + a)(\p_x + b) - ab - b_y + c
\end{array}\right.
\end{equation}

and this presentation is  a first step of Darboux-Laplace
transformations used for construction of  solution to
non-factorizable linear partial differential equations in two
independent variables.\\

To generalize  construction of invariants to the case of an
arbitrary order operator, a factorization algorithm for such an
operator is needed. Recently two factorization algorithms for
arbitrary order bivariate LPDOs have been published. In \cite{gs},
the factorization algorithm called {\it Hensel descent} is
presented, which is a modification of well-known Hensel lifting
algorithm (see, for instance, \cite{Kalt}) and which allows to
factorize operators from the ring $D=\Rational(x,y)[\px, \py].$ In
\cite{bk2005}, {\it absolute} factorization is presented for
operators whose coefficients are arbitrary smooth functions of two
variables. The idea to connect the results produced by this method
(called now BK-factorization) and Darboux-Laplace transformations
has been briefly discussed in \cite{k2005}. In present paper we
construct general invariants using BK-factorization procedure
which can be described as follows. Factorization problem is
regarded in the form

\begin{equation}\label{An}
A_n=\sum_{j+k\le n}a_{jk}\partial_x^j\partial_y^k =(p_1\partial_x
+ p_2 \partial_y + p_3)\Big(\sum_{j+k <
n}p_{jk}\partial_x^k\partial_y^j\Big)
\end{equation}
with  assumption that  $a_{n0} \neq 0$ (locally) and can be taken
as $a_{n0} = 1$ without loss of generality. It is shown that for
any simple root $\o$  of characteristic polynomial

$$
A(\o)=\sum^n_{k=0}a_{n-k,k}\o^{n-k},
$$

with $n \geq 2$, there exists unique factorization of the form
(\ref{An}):

\begin{equation}\label{factor}
A=LB+C, \qquad L=\pa_x-\o\pa_y+p, \qquad
C=\sum^{n-2}_{j=1}c_j\pa_y^j.
\end{equation}

and the conditions of factorization are written out explicitly. In
Section 2 we use these results for construction of general
invariants of an arbitrary LPDO and study some of their properties
while in Section 3 the whole hierarchy of general invariants is
described. In Section 4 a few instructive examples are given. In
particular, it is shown that both Laplace invariants can be
constructed as a simple particular case of general invariants. In
Section 5 brief discussion of the results obtained is presented.

\section{General invariants and semi-invariants}

The conditions of factorization produced by BK-factorization
have following form:\\

for second order operator
$$ A_2=\sum_{j+k\le2}a_{jk}\px^j\py^k
=(p_1\px+p_2\py+p_3)(p_4\px+p_5\py+p_6)$$

the only factorization condition is

\be\label{cond2} a_{00} = L(p_6)+p_3p_6, \ee

for third order operator
$$
A_3=\sum_{j+k\le3}a_{jk}\partial_x^j\partial_y^k
=(p_1\px+p_2\py+p_3)(p_4 \p_x^2 +p_5 \px\py  + p_6 \py^2 + p_7 \px
+ p_8 \py + p_9)$$
 there are two factorization conditions:

\be \label{cond3}
\begin{cases}
a_{01} = \ L(p_8)+p_3p_8+p_2p_9\\
a_{00} = \ L(p_9)+p_3p_9
\end{cases}
\ee

and so on for arbitrary $n$. Now we can give explicit
representation of a LPDO
 whose characteristic polynomial has at least one  simple root,
 in the form of factorization with  remainders:

$$
 A_2
=(p_1\px+p_2\py+p_3)(p_4\px+p_5\py+p_6)-l_2
$$
with remainder
$
l_2=  a_{00} - L(p_6)+p_3p_6;
$
$$ A_{3}=(p_1\px+p_2\py+p_3)(p_4 \p_x^2 +p_5 \px\py  + p_6 \py^2 +
p_7 \px + p_8 \py + p_9)-l_3\py-l_{31}, $$ with  two remainders
$l_3= a_{01}- (p_1\px+p_2\py+p_3)p_8-p_2p_9,
 \ \ \mbox{and} \ \
 l_{31}=a_{00}-(p_1\px+p_2\py+p_3)p_9;
 $
and so on for arbitrary order of LPDO.\\

\paragraph{Definition} The
operators $A$, $\tilde A$ are called equivalent if there is a
gauge transformation that takes one to the other:
$$
\tilde Ag= e^{-\vp}A(e^{\vp}g)\equiv A_\vp g.
$$
Note that
$$
(\pa_x)_\vp=\pa_x+\vp_x,\quad (\pa_y)_\vp=\pa_y+\vp_y
$$
and a gauge transformation is an algebra automorphism.  Therefore
$A$ and $A_\vp$ have the same characteristic polynomial and the
factorization \eqref{factor} carries over:
$$
A_\vp=L_\vp B_\vp +C_\vp.
$$
It follows that both the characteristic polynomial and the leading
nonzero coefficient of the remainder term are invariants. Number
of remainders varies for operators of different orders, and also
their properties are different. In order to demonstrate it, let us
formulate the following

\paragraph{Lemma 1}
 For an operator of order 2, its remainder $l_2$ is
its {\bf invariant} under
 the equivalence transformation, i.e.
$$ \tilde{l}_{2}=l_{2}. $$ For an operator of order 3, its remainder
${l}_{3}$ is its {\bf invariant}, i.e. $$\tilde{l}_{3}=l_{3},$$
while remainder ${l}_{31}$ changes its form as follows:
 $$\tilde{l}_{31}=l_{31}+l_{3}\varphi_y. $$

$\blacktriangleright$ Introducing notations $ \ \
A_{2a}=A_{2p}-l_2, \ \ A_{3a}=A_{3p}-l_{3}\py-l_{31}, \ \ f=e^\vp,
\ \ $ we get
$$
\tilde{A}_{2a}=f^{-1}A_{2a}\circ f= f^{-1} (A_{2p}-l_2)\circ
f=\tilde{A}_{2p}-f^{-1}(l_2)\circ f=\tilde{A}_{2p}-l_2.
$$
and
$$
\tilde{A}_{3a}=f^{-1}A_{3a}\circ f= f^{-1}
(A_{3p}-l_{3}\py-l_{31})\circ
f=\tilde{A}_{3p}-f^{-1}(l_{31}+l_{3}\py)\circ
f=\tilde{A}_{3p}-l_{3}\py-l_{3}\varphi_y-l_{31}.\qed
$$

\paragraph{\bf Corollary 1: }
If $l_{3}=0$, then $l_{31}$ becomes invariant.\\

That is the reason why we call $l_{31}$ further {\bf
semi-invariant}.

\paragraph{\bf Corollary 2: }
If $l_{3} \neq 0$, it is always possible to choose some function
$f: \quad \tilde{l}_{31}= l_{3}\varphi_y+l_{31}=0$.\\

Note that for second order operator, if its invariant
 $l_2=0$ then operator is factorizable while for third order
 operator two its invariants have to be equal to zero,
 $l_{3}=l_{31}=0$. On the other hand, if operator of third order
 is not factorizable we can always regard it as an operator with
only one non-zero invariant. Of course, all this is true for each
simple root of characteristic polynomial, so that one expression,
say, for $l_{3}$, will generate three invariants in
case of three simple roots of corresponding polynomial.\\

\section{Hierarchy of invariants}

As it was shown above, every general invariant is a function of a
simple root $\o$ of the characteristic polynomial and each simple
root provides one invariant. It means that for operator of order
$n$ we can get no more than $n$ different invariants. Recollecting
that BK-factorization in this case gives us one first order
operator and one operator of order $n-1$, let us put now following
question: are general invariants of operator of order $n-1$ also
invariants of
corresponding operator of order $n$?\\

Let regard, for instance, operator of order 3
$$
{A}_{3a}={A}_{1}{A}_{2a}-l_{3}\py-l_{31}=
{A}_{1}({A}_{2p}-l_2)-l_{3}\py-l_{31}={A}_{1}{A}_{2p}-l_2{A}_{1}-l_{3}\py-l_{31}$$
and
$$
\tilde{A}_{3a}={A}_{1}{A}_{2p}-l_2{A}_{1}-l_{3}\py-l_{31}=
\tilde{A}_{1}\tilde{A}_{2p}-l_2\tilde{A}_{1}-l_{3}\py-\tilde{l}_{31},
$$
i.e. $l_2$ is also invariant of operator ${A}_{3a}$. Let us notice
that general invariant $l_3=l_3(\o^{(3)})$ is a function of a
simple root $\o^{(3)}$ of the polynomial
$$
\mathcal{P}_3(z)= a_{30}z^3+a_{21}z^2+a_{12}z+a_{03}
$$
while general invariant $l_2=l_2(\o^{(2)})$ is a function of a
simple root $\o^{(2)}$ of the polynomial
$$
\mathcal{R}_2(z)=p_4 z^2 +p_5 z  + p_6
$$
with $p_4, \ p_5, \ p_6$ given explicitly for $\o=\o^{(3)}$. In
case of all simple roots of both polynomials $\mathcal{P}_3(z)$
and $\mathcal{R}_2(z)$, one will get maximal number of invariants,
namely 6 general invariants. Repeating the procedure for an
operator of order $n$, we get maximally $n!$ general invariants.
In this way for operator of arbitrary order $n$ we can construct
the hierarchy of its general invariants
$$
l_n, l_{n-1}, ..., l_2
$$
and their explicit form is given by BK-factorization. As to
semi-invariants, notice that an operator of arbitrary order $n$
can always be rewritten in the form of factorization with
remainder of the form
$$
l_n\px^k+l_{n,1}\px^{k-1}+...+l_{n,k-1}, \ \ k<n
$$
and exact expressions for all $l_i$ are provided by
BK-factorization procedure. The same reasoning as above will show
immediately that $l_n$ is always  general invariant, and each
$l_{n,k-i_0}$ is $i_0$-th semi-invariant, i.e. it becomes
invariant in case if $l_{n,k-i}=0, \ \forall i<i_0$.\\

\paragraph{Example 1} Let us regard a third order hyperbolic operator in
the form

\be \label{ex4} C=a_{30}\p_x^3 + a_{21}\p_x^2 \py + a_{12}\px
\py^2 +a_{03}\p y^3+\mbox{terms of lower order} \ee

 with constant high order coefficients, i.e.
$a_{ij}=\const \ \forall \ i+j=3$ and all roots of characteristic
polynomial
$$
a_{30}\o^3+a_{21}\o^2+a_{12}\o+a_{03}=\mathcal{P}_3(\o)
$$
are simple and real.  Then we can construct three simple
independent general invariants in following way. Notice first that
in this case high terms of (\ref{ex4}) can be written in the form
$$
(\a_1 \px + \b_1 \py)(\a_2 \px + \b_2 \py)(\a_3 \px + \b_3 \py)
$$
 for all non-proportional $\a_j, \b_i$ and after appropriate change
 of variables this expression can easily be reduced to
 $
\px \py (\px+ \py).
 $
Let us introduce notations $\pa_1=\pa_x,\ \ \pa_2=\pa_y,\ \
\pa_3=\pa_1+\pa_2=\p_t,$ then all terms of the third and second
order can be written out as
$$C_{ijk}=(\pa_i+a_i)(\pa_j+a_j)(\pa_k+a_k)=\pa_i\pa_j\pa_k+
a_k\pa_i\pa_j+a_j\pa_i\pa_k+a_i\pa_j\pa_k+$$
$$+
(\pa_j+a_j)(a_k)\pa_i+(\pa_i+a_i)(a_k)\pa_j+(\pa_i+a_i)(a_j)\pa_k+
(\pa_i+a_i)(\pa_j+a_j)(a_k)$$
 with $$a_{20}=a_2, \
a_{02}=a_1, \ a_{11}=a_1+a_2+a_3$$ and $c_{ijk}=C-C_{ijk}$ is an
operator of the first order which can be written out explicitly.
As it was shown above, coefficients of $c_{ijk}$ in front of first
derivatives  are (semi)invariants and therefore, any linear
combination of invariants is an invariant itself. Direct
calculation gives us three simplest general invariants of the
initial operator $C$:
$$l_{21}=a_{2,x}-a_{1,y}, \ l_{32}=a_{3,y}-a_{2,t}, \ l_{31}=a_{3,x}-a_{1,t}.$$

\paragraph{Lemma 2} General invariants $l_{21}, \ l_{32}, \ l_{31}$
are all equal to zero {\bf iff} operator $C$ is equivalent to an
operator
 \be \label{Prop}L=\pa_1\pa_2\pa_3+b_1\pa_1+b_2\pa_2+c,\ee
i.e. $\exists \ \mbox{function } \ f: \ \   f^{-1}C \circ f =L.$\\

 $\blacktriangleright$ Obviously
 $$
f^{-1}(\px \py \p_t )\circ f= (\px +(\log f)_x)(\py +(\log
f)_y)(\p_t +(\log f)_t)
 $$
for any smooth function $f$. Notice that it is the form of an
operator $C_{ijk}$ and introduce a function $f$ such that

$$
a_1=(\log f)_x, \ \ a_2= (\log f)_y, \ \ a_3= (\log f)_t.
$$

This system of equations on $f$ is over-determined
 and it has  solution $f_0$ {\bf iff}
\be \label{sys} a_{2,x}-a_{1,y}=0, \ a_{3,y}-a_{2,t}=0, \
a_{3,x}-a_{1,t}=0,\ee i.e. $l_{21}= l_{32}= l_{31}=0.$

$\blacktriangleleft$ Indeed, if $C$ is equivalent to (\ref{Prop}),
then $a_{20}=a_{02}=a_{11}=0$ and obviously
$l_{21}=l_{32}=l_{31}=0.$ \qed \\

Note that the statement of the Lemma 2 is  weaker than that of
Theorem 1.

\paragraph{Example 2}
First two equations of (\ref{sys}) determine solution uniquely, up
to a constant, and therefore determine
$$
\partial_3 f=(\partial_1+\partial_2) f =a_1+a_2.
$$
However the compatibility conditions are satisfied if
$\partial_2a_1=\partial_1a_2$ and $a_3=a_1+a_2+c$, $c$ constant.
In general, the compatibility conditions only tell that two
equations of (\ref{sys}) are satisfied, but the third may be off
by a constant. For instance, two constant coefficient operators
$$
\partial_x\partial_y(\partial_x+\partial_y),\qquad
\partial_x \partial_y(\partial_x+\partial_y+1)
$$
have the same general invariants (all zero) but are not
equivalent.

\section{Further examples}

\paragraph{Example 3} Let us regard hyperbolic operator in the form
\begin{equation}\label{ex1}
\partial _{xx} - \partial_{y y} + a_{10} \partial_{ x} +
a_{01} \partial_{ y} + a_{00},
\end{equation}
 i.e. $a_{20}=1, a_{11}=0, a_{02}=-1$ and $\o=\pm 1, \ \mathcal{L}=\px-\o \py.$
Then $l_2$ takes form
$$
l_2=a_{00}-\mathcal{L}(\frac{\o a_{10}-a_{01}}{2\o})-\frac{\o
a_{10}-a_{01}}{2\o} \frac{\o a_{10}+a_{01}}{2\o}
$$
which yields, for instance for the root $\o=1$, to
$$
l_2=a_{00}-\mathcal{L}(\frac{a_{10}-a_{01}}{2})-\frac{
a_{10}^2-a_{01}^2}{4}=a_{00}-(\px-
\py)(\frac{a_{10}-a_{01}}{2})-\frac{ a_{10}^2-a_{01}^2}{4}
$$
and after obvious change of variables in (\ref{ex1}) we get
finally first Laplace invariant $\hat{a}$
$$
l_2= c-\p_{\tilde{x}}a -ab = \hat{a},
$$
where
$$
a=\frac{a_{10}-a_{01}}{2}, \ \ b=\frac{a_{10}+a_{01}}{2}, \ \
c=a_{00}.
$$
Choice of the second root, $\o=-1$, gives us the second Laplace
invariant $\hat{b}$, i.e. Laplace invariants are particular cases
of the general invariant so that each Laplace invariant
corresponds to a special choice of $\o$.

\paragraph{Example 4}
Let us proceed analogously with an elliptic operator
\begin{equation}\label{ex2}
\partial _{xx} + \partial_{y y} + a_{10} \partial_{ x} +
a_{01} \partial_{ y} + a_{00},
\end{equation}
then $\o=\pm i, \ \mathcal{L}= \px-\o \py$ and
$$
l_2=a_{00}+(\px \mp i \py)(\frac{\pm a_{10}+a_{01}i}{2})+i\frac{
a_{10}^2+a_{01}^2}{4}
$$
where choice of upper signs corresponds to the choice of the root
$\o=i$ and choice of lower signs corresponds to $\o=-i$.

\paragraph{Example 5}
Now let us regard a third order operator

\be\label{ex3} B=\p_x^2 \py + \px \py^2 +a_{11}\px\py
+a_{10}\px+a_{01}\py+a_{00}, \ee

with $a_{30}=a_{03}=a_{20}=a_{02}=0,\ \ a_{21}=a_{12}=1$ and all
other coefficients are functions of $x,y$. Then (semi)invariants
$$l_{3}=\px a_{11}-a_{01} \ \  \mbox{and} \ \
l_{31}=\px a_{10}-a_{00}$$
have very simple forms and gives us
immediately a lot of information about the properties of operators
of the form (\ref{ex3}), for instance, these operators are
factorizable, i.e. has zero invariants $l_3=l_{31}=0$, {\bf iff}
$$a_{11}=\int a_{01}dx + f_1(y), \ \ a_{10}= \int a_{00}dx +
f_2(y)
$$ with two arbitrary functions on $y$, $f_1(y)$ and $f_2(y)$.

\section{Summary}

Now, that construction of Laplace invariants has been generalized
to the case of arbitrary order and arbitrary type of LPDO, the
next obvious step would be to generalize of Darboux-Laplace
transformations to this class class of operators, to construct
generalized Todda lattice, etc. It would require a deeper study of
 intrinsic properties of general invariants. For instance,
beginning with operator of order 4, their maximal number  is
bigger then number of coefficients of a given operator,
$$\frac{(n+1)(n+2)}{2} < n! \ \ \forall n>3.$$ It means that
general invariants are dependent on each other and it will be a
challenging task to extract the subset of independent general
invariants, i.e. basis in the finite space of general invariants.
Another important step
would be to extract a subset providing full analog of Theorem 1
for arbitrary order operators.\\

Already in the case of three variables, the factorization problem
of a corresponding operator and also constructing of its
invariants becomes more complicated, even for constant
coefficients. The reason of it is that in bivariate case we needed
just to factorize leading term polynomial which is always possible
over $\Complex$. It is not the case for more then 2 independent
variables where a counter-example is easily to find:

\paragraph{Example 6}  A polynomial
$$
x^3+y^3+z^3-3xyz=(x+y+z)(x^2+y^2+z^2-xy-xz-zy)
$$
is factorizable but it is easy to prove that $ x^3+y^3+z^3$ is not
divisible by a linear polynomial $$\a x+ \b y+ \g z + \d$$ for any
complex coefficients $\a,\ \b,\ \g, \ \d$. Thus, some non-trivial
conditions are to be found for factorization of polynomials in
more then two variables.

\section*{Acknowledgements}
 Author
acknowledges support of the Austrian Science Foundation (FWF)
under projects SFB F013/F1304. Author is  grateful to Prof. A.
Shabat  for useful discussions during preparing of this paper and
also to Prof. S. Tsarev who noticed in particular that hierarchy
of invariants constructed above can be regarded as analog of
classical Todda lattice. Author is also very much obliged to an
anonymous reviewer and  Prof. R. Beals for their valuable help
during preparing of the last version of this paper.

\end{document}